\documentclass[twocolumn,english,showpacs,superscriptaddress,floatfix,nofootinbib]{revtex4}
\usepackage[T1]{fontenc}
\usepackage[latin9]{inputenc}
\usepackage{amsmath}
\usepackage{amssymb}
\usepackage{graphicx}
\usepackage{esint}

\makeatletter
\@ifundefined{textcolor}{}
{%
 \definecolor{BLACK}{gray}{0}
 \definecolor{WHITE}{gray}{1}
 \definecolor{RED}{rgb}{1,0,0}
 \definecolor{GREEN}{rgb}{0,1,0}
 \definecolor{BLUE}{rgb}{0,0,1}
 \definecolor{CYAN}{cmyk}{1,0,0,0}
 \definecolor{MAGENTA}{cmyk}{0,1,0,0}
 \definecolor{YELLOW}{cmyk}{0,0,1,0}
}

\usepackage[active]{srcltx}\usepackage{float}\usepackage{epsfig}\usepackage{babel}
\usepackage{babel}
\usepackage{babel}
\usepackage{babel}
\usepackage{babel}
\usepackage{babel}
\usepackage{babel}
\usepackage{babel}
\usepackage{babel}
\usepackage{babel}
\usepackage{babel}
\usepackage{babel}
\usepackage{babel}
\usepackage{babel}
\usepackage{babel}
\usepackage{babel}
\usepackage{babel}
\usepackage{babel}

\@ifundefined{textcolor}{}{
 \definecolor{BLACK}{gray}{0}
 \definecolor{WHITE}{gray}{1}
 \definecolor{RED}{rgb}{1,0,0}
 \definecolor{GREEN}{rgb}{0,1,0}
 \definecolor{BLUE}{rgb}{0,0,1}
 \definecolor{CYAN}{cmyk}{1,0,0,0}
 \definecolor{MAGENTA}{cmyk}{0,1,0,0}
 \definecolor{YELLOW}{cmyk}{0,0,1,0}
}
\@ifundefined{textcolor}{}{
 \definecolor{BLACK}{gray}{0}
 \definecolor{WHITE}{gray}{1}
 \definecolor{RED}{rgb}{1,0,0}
 \definecolor{GREEN}{rgb}{0,1,0}
 \definecolor{BLUE}{rgb}{0,0,1}
 \definecolor{CYAN}{cmyk}{1,0,0,0}
 \definecolor{MAGENTA}{cmyk}{0,1,0,0}
 \definecolor{YELLOW}{cmyk}{0,0,1,0}
}

\DeclareFontEncoding{LGR}{}{}
\DeclareTextSymbol{\~}{LGR}{126}
\@ifundefined{textcolor}{}{
 \definecolor{BLACK}{gray}{0}
 \definecolor{WHITE}{gray}{1}
 \definecolor{RED}{rgb}{1,0,0}
 \definecolor{GREEN}{rgb}{0,1,0}
 \definecolor{BLUE}{rgb}{0,0,1}
 \definecolor{CYAN}{cmyk}{1,0,0,0}
 \definecolor{MAGENTA}{cmyk}{0,1,0,0}
 \definecolor{YELLOW}{cmyk}{0,0,1,0}
}

\usepackage{babel}

\makeatother

\usepackage{babel}
\begin{document}

\title{Conductance spectroscopy in ferromanget/superconductor hybrids}

\author{T.~Yu.~Karminskaya}

\affiliation{Skobeltsyn Institute of Nuclear Physics, Lomonosov Moscow State University,
Leninskie gory, Moscow 119991, Russian Federation}

\author{M.~Yu.~Kupriyanov}

\affiliation{Skobeltsyn Institute of Nuclear Physics, Lomonosov Moscow State University,
Leninskie gory, Moscow 119991, Russian Federation}

\author{S.L. Prischepa}

\affiliation{Belarusian State University of Informatics and
Radio Electronics, Minsk, Belarus}

\author{A.~A.~Golubov}

\affiliation{Faculty of Science and Technology and MESA+ Institute for Nanotechnology,
University of Twente, 7500 AE Enschede, The Netherlands}

\begin{abstract}
We present theoretical model for the proximity effect in F-SFF-F structures (F is ferromagnet, S is superconductor)
with non-collinear magnetizations vectors in the F-layers and with arbitrary magnitudes of exchange
fields. Electrical conductance of these structures is analyzed within the Keldysh-Usadel formalism in the diffusive regime
as a function of a misorientation angle between magnetizations of the F-layers and
transparencies of SF and FF interfaces. We show that long-range triplet
superconducting correlations manifest themselves either as a zero-bias peak in the case of perfect transparency of FF interface,
or as a two-peak structure in the case of finite transparency. The predicted features may serve as a
diagnostic tool for characterization of interfaces in superconducting hybrid structures.
\end{abstract}
\maketitle

Investigation of superconducting correlations
in superconductor (S) - ferromagnetic (F) hybrid structures is currently a subject
of active interest. Quite a number of remarkable phenomena were predicted theoretically in these structures \cite{Bulaev}-\cite{Kawabata}
and experimentally verified \cite{Ryazanov2001}-\cite{exp13}.
Moreover, potential applications of SF-based devices as memory elements
in superconducting computers were recently proposed \cite{Herr1},\cite{Herr2}, \cite{AIRPA}.

There are several types of spin valve devices that potentially can
be used as memory elements \cite{exp3}-\cite{exp13}, \cite{Ryazanov}-\cite{Bakurskiy2}.
Among them, only the structure proposed in \cite{Ryazanov}-\cite{Bakurskiy2}
can operate in magnetic fields that do not exceed a few tens of oersted.
In these devices there is only one ferromagnetic film, so only 
short-range spin triplet correlations are present in such structures.
Contrary to that, in other spin valve realizations \cite{exp3}-\cite{exp13}
there are several magnetic layers. In these structures, deviations of relative magnetizations
of ferromagnetic films from collinear to non-collinear one leads
to generation of long-range triplet superconducting correlations.
This process is accompanied by either suppression of the
critical temperature \cite{exp12}, \cite{exp13}, $T_{C},$ or by changing
sign of the supercurrent in the triplet pairing channel
\cite{exp3}-\cite{exp11}. In both cases an implementation of these
effects requires an application of an external magnetic field of
the order of $10^{3}$ Oe. It is obvious that such large magnetic fields
cannot be easily combined with the RSFQ circuits.

An alternative is to use structures with long ferromagnetic films
suggested in \cite{Karminskaya3}. The decay length of long range
triplet superconducting component is insensitive to the magnitude of exchange
field, therefore these correlations penetrate into ferromagnetic material at longer
distances. Such correlations can be observed in long ferromagnetic
wires in the parts where singlet and short-range triplet
correlations are suppressed due to their fast decay in space
intrinsic to materials having large magnitude of the exchange
energy. The long-range 
triplet correlations can survive in long ferromagnetic films attached
to SFF structures in which 
long-range triplet pairing can nucleate. Presence of these 
 triplet correlations changes density of states (DoS) in the film, so it must be also
accompanied by changes in the conductance of this film due to proximity
effect. It provides an opportunity for realization of a spin valve
in which the external magnetic field controls the magnitude of conductance
of ferromagnetic film.

To evaluate the magnitude of this effect we have analyzed a simple
model problem below. Namely, we investigate correlations in F1-SF1F2-F1
structures (see Fig. 1) that represent a long thin ferromagnetic wire
F1 with the length $2L+d_{s}.$ It connects two massive normal electrodes.
In its middle part the F1 layer is in contact with thick superconducting
film S located on the top of the wire and thin ferromagnetic film
F2 placed on the bottom of the wire (the lengths, $d_{s},$ of S and
F2 films are identical). Magnetization vector of the long ferromagnetic
wire is constant and it is directed along F1F2 interface. Magnetization
vector of the short ferromagnetic film F2 is declined from the first
vector on an angle $\alpha$ thus providing conditions for realization
of
long-range triplet superconducting correlations in the structure.
Here we present results of calculations for DoS and differential conductance
along F1 film in F1-SF1F2-F1 structure with noncollinear magnetizations
of ferromagnetic layers that also differs with values of exchange
energies. The calculations are done in the diffusive limit
in the framework of the Usadel equations for both linear and nonlinear
cases. We present differential conductance of the F1 film as a function
of angle $\alpha$. We show that maximum value of differential conductance
is achieved not at $\alpha=\pi/2$ (as it was found in \cite{BVE1}
and \cite{Vasenko} for out of plane geometry for normal current injection),
but at some intermediate angle that depends on difference between
the values of exchange energy of F films. We also investigate influence
of suppression parameter on F1F2 interface on the shape of differential
conductance.

\begin{figure}
\centerline{\includegraphics[scale=0.15]{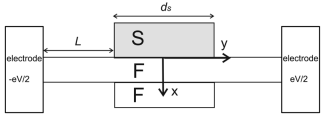}} \caption{Geometry of the considered F1-SF1F2-F1 structure.}
\end{figure}

We will discuss properties of the structure 
in the frame of Usadel equations that can be written as:
\begin{equation}
D\nabla(G\nabla G)+i\varepsilon\lbrack\hat{k}_{0}\hat{\tau}_{3}\hat{\sigma}_{0},G]-i[\hat{k}_{0}\overline{h},G]=0,\label{eq:us_g}
\end{equation}
where parameter $\overline{h}=h_{2}(\hat{\tau}_{3}\hat{\sigma}_{3}\cos\alpha+\hat{\tau}_{0}\hat{\sigma}_{2}\sin\alpha)$
for F2 layer and $\overline{h}=h_{1}\hat{\tau}_{3}\hat{\sigma}_{3}$
for F1 film, ($\hat{k},\hat{\:\tau,\:}\hat{\sigma}$ are $2\times2$
Pauli matrices), $h_{1}$ and $h_{2}$ are normalized on $\pi T_{C}$
exchange energies of upper and lower F films, respectively, $D$ is
the diffusion coefficient, $\varepsilon$ is a quasiparticle energy.
Green's function $G$ is $8\times8$ matrix
\[
G=\left(\begin{array}{cc}
G^{R} & G^{K}\\
0 & G^{A}
\end{array}\right).
\]
Here $G^{R},G^{A},G^{K}$ are retarded, advanced and Keldysh Green's
functions, correspondingly. 
The elements of matrix $G^{K}$ can be connected by distribution function
$f=f_{L}\hat{\tau}_{0}\hat{\sigma}_{0}+f_{T}\hat{\tau}_{3}\hat{\sigma}_{0}$
(which is $4\times4$ matrix):

\begin{equation}
G^{K}=G^{R}f-fG^{A}.\label{eq:Gk}
\end{equation}
In thermodynamic equilibrium, that in our case is achieved in the
electrodes (see Fig.1), the distribution function is expressed by

\begin{equation}
f_{L,T}=\frac{1}{2}[\tanh\left(\frac{\varepsilon+eV}{2T}\right)\pm\tanh\left(\frac{\varepsilon-eV}{2T}\right)],\label{eq:equil_dis_f}
\end{equation}
where $V$ is an applied voltage (Fig. 1). Consequently, it is sufficient
to determine only retarded function. This function can be represented
as
\[
G^{R}=\hat{\tau}_{3}g+\hat{\tau}_{0}g_{t}+i\hat{\tau}_{2}f+\hat{\tau}_{1}f_{t},
\]
\[
f=\hat{\sigma}_{3}f_{3}+\hat{\sigma}_{0}f_{0},f_{t}=\hat{\sigma}_{1}f_{1},
\]
\[
g=\hat{\sigma}_{0}g_{0}+\hat{\sigma}_{3}g_{3},:g_{t}=\hat{\sigma}_{2}g_{2},
\]
where $f_{3},f_{0},f_{1}$ are condensate Green's functions describing
singlet, short-range and long-range triplet correlations; $g_{0},g_{3},g_{2}$
are normal Green functions, respectively. With these definitions normalization
condition\emph{\ $G^{2}=1$ }transforms to
\begin{equation}
g_{0}^{2}+g_{3}^{2}+g_{2}^{2}-f_{3}^{2}-f_{0}^{2}+f_{1}^{2}=1,\label{eq:nc}
\end{equation}
\[
f_{1}f_{3}+g_{2}g_{0}=0.
\]
Equation (\ref{eq:us_g}) must be supplemented by boundary conditions
matching Green's functions across the interfaces
\begin{equation}
\gamma_{B}G_{l}\frac{\partial}{\partial x}G_{l}=\pm\lbrack G_{l},G{}_{r}],\label{eq:bc_1}
\end{equation}
\begin{equation}
\gamma G_{l}\frac{\partial}{\partial x}G_{l}=G_{r}\frac{\partial}{\partial x}G_{r}.\label{eq:bc_2}
\end{equation}
At $y=\pm(L+d_{s}/2)$

\begin{equation}
G_{l}=0,\label{eq:bc_3}
\end{equation}
while electron energy distribution functions are equal to their equilibrium
values (\ref{eq:equil_dis_f}). Here the indices, $l,$ and, $r,$
refer to the upper or lower layer with respect to the SF1 and F1F2
boundaries.

Transport properties of both F1F2 and F1S interfaces are characterized
by the interface parameters
\begin{equation}
\gamma=\frac{\rho_{S}\xi_{S}}{\rho_{F}\xi_{F}},\quad\gamma_{B}=\frac{R_{BF}\mathcal{A}_{B}}{\rho_{F}\xi_{F}},\quad\gamma_{BS}=\frac{R_{BS}\mathcal{A}_{B}}{\rho_{F}\xi_{F}}.\label{gammas}
\end{equation}
Here $R_{BF},$ $R_{BS}$ and $\mathcal{A}_{B}$ are the resistances
and area of the F1F2 and F1S interfaces, $\xi_{S,F}=(D_{S,F}/2\pi T_{C})^{1/2}$
and, $D_{S,F},$ are the decay lengths and diffusion coefficients
of S and F materials, while $\rho_{S}$ and $\rho_{F}$ are their
resistivities.

To simplify the problem we assume below that normal state resistivities
and coherence lengths of ferromagnetic films are identical ($\gamma=1$
at F1F2 interface), ferromagnetic films are thin. The second assumption
allows us to transfer solution of the problem (\ref{eq:us_g})-(\ref{eq:bc_2})
to a one-dimensional one (see \cite{Karminskaya1}-\cite{Karminskaya2}
for the details). We also assume that suppression parameters at F1S
interface satisfy the condition $\gamma\ll\xi_{F}/(d_{F}max(1,\sqrt{h_{1,2}}))+\gamma_{BS}$
allowing to ignore the suppression of superconductivity in S electrode.
We assume further, that the length, $L,$ is smaller compare to the
characteristic lengths of inelastic scattering inside the F1 layer.

Under the above assumptions it is possible to derive from (\ref{eq:us_g})
the equation for electron energy distribution for F1-SF1F2-F1 structure
in the form of diffusion equation
\[
\frac{\partial}{\partial x}(M\frac{\partial}{\partial x}f_{T})=0,
\]
\[
M=[Reg_{0}]^{2}+[Reg_{3}]^{2}+[Reg_{2}]^{2}+[Imf_{3}]^{2}+[Imf_{0}]^{2}+[Ref_{1}]^{2}.
\]
Integrating this equation with boundary conditions (\ref{eq:bc_3})
we get expression for distribution function $f_{T}$ and from the
general expression for current
\begin{equation}
I=\frac{1}{2R}\intop Tr\sigma_{0}\tau_{3}(G^{R}\frac{\partial}{\partial y}G^{K}+G^{K}\frac{\partial}{\partial y}G^{A})d\varepsilon
\end{equation}
and (\ref{eq:Gk}) we arrive at
\begin{equation}
I=\frac{d}{R}\intop_{0}^{\infty}\frac{2f_{T}}{\intop_{-d}^{d}\frac{dy}{M}}d\varepsilon,\label{eq:current}
\end{equation}
where $d=L+d_{s}/2.$

In the limit of zero temperature from (\ref{eq:current}) for normalized
differential conductance $\sigma(V)=RdI/dV$ of long ferromagnetic
film in the direction along F1F2 interface we have
\begin{equation}
\sigma=\frac{2d}{\intop_{-d}^{d}\frac{dy}{[Reg_{0}]^{2}+[Reg_{3}]^{2}+[Reg_{2}]^{2}+[Imf_{3}]^{2}+[Imf_{0}]^{2}+[Ref_{1}]^{2}}}.\label{eq:cond}
\end{equation}
\qquad{}\qquad{}

Expression (\ref{eq:cond}) can be simplified in the linearized case
when suppression parameter $\gamma_{BS}$ is large enough. In this
limit superconductivity induced into F1 wire is small and in the zero
approximation $g_{0}=1,$ $g_{3,2}=0.$ \ Taking into account normalization
condition (\ref{eq:nc}), we can express $g_{0}$ through the other
functions in the next approximation and transforms expression (\ref{eq:cond})
to:
\begin{equation}
\sigma=\frac{2d}{\intop_{-d}^{d}\frac{dy}{1+[Ref_{3}]^{2}+[Ref_{0}]^{2}+[Imf_{1}]^{2}}}.\label{eq:condlin}
\end{equation}

Note, that expression for differential conductance similar to Eq.
(\ref{eq:condlin}) was obtained for SNN' structure $(Ref_{0}=0,Imf_{1}=0)$
in \cite{Volkov}. Conductance in superconducting hybrids in a similar T-shaped geometry was further studied
in \cite{Nazarov,Golubov-1} (see review in  \cite{Belzig}) and in \cite{Asano2007}. However, in the structures
considered in \cite{Nazarov,Golubov-1,Belzig} no ferromagnetic layers were attached
and therefore there were no odd superconducting correlations. In the
structure considered in \cite{Asano2007} there also were no ferromagnetic
layers, but odd triplet correlations were generated due to proximity effect between a p-wave superconductor
and a diffusive N-layer.

We start our analysis of processes in F1-SF1F2-F1 structure by considering
the case of a transparent F1F2 interface ($\gamma_{B}=0.$) This limiting
case is simple. However, it reveals the main effects without any distortion
due to the influence of the F1F2 interface. In the linearized case
analytical solutions of the problem ((\ref{eq:us_g})-(\ref{eq:bc_3}))
for condensate functions $f_{3},f_{0},f_{1}$ in free part of upper
F film can be easily derived:
\begin{equation}
f_{1}=\Gamma\frac{h_{2}\sin(\alpha)\frac{\sinh(q(d-y))}{\sinh(q(d-d_{S}/2))}}{h_{1}^{2}+h_{2}^{2}+2h_{1}h_{2}\cos(\alpha)-4\varepsilon^{2}},\label{eqff0}
\end{equation}
\begin{equation}
f_{0}=i\frac{\Gamma}{2}\frac{\sum_{j=1,2}\frac{\left[(-1)^{j+1}2\varepsilon+h_{1}+h_{2}\cos(\alpha)\right]\sinh(q_{j}(d-y))}{\sinh(q_{j}(d-d_{S}/2))}}{h_{1}^{2}+h_{2}^{2}+2h_{1}h_{2}\cos(\alpha)-4\varepsilon^{2}},\label{eq:ff}
\end{equation}
\begin{equation}
f_{3}=i\frac{\Gamma}{2}\frac{\sum_{j=1,2}\frac{\left[2\varepsilon+(-1)^{j+1}(h_{1}+h_{2}\cos(\alpha))\right]\sinh(q_{j}(d-y))}{\sinh(q_{j}(d-d_{S}/2))}}{h_{1}^{2}+h_{2}^{2}+2h_{1}h_{2}\cos(\alpha)-4\varepsilon^{2}}.\label{eq:ff-1}
\end{equation}
In these expressions $q=\sqrt{-i\varepsilon}$, $q_{1,2}=\sqrt{-i\varepsilon\pm ih_{1,2}}$,
$\Gamma=\Delta/(\gamma_{BMS}\sqrt{\varepsilon^{2}-\Delta^{2}})$,
$\Delta$ is the modulus of the order parameter of superconductor,
$\gamma_{BMS}=\gamma_{BS}d_{F}/\xi_{F}$ . It is necessary to note
that to get (\ref{eqff0})-(\ref{eq:ff-1}) we also neglect suppression
of Green's functions in the part of F1 film located under superconductor
due to proximity effect and use the rigid boundary conditions at $x=\pm d_{s}/2.$
\begin{figure}
\centerline{\includegraphics[scale=0.9]{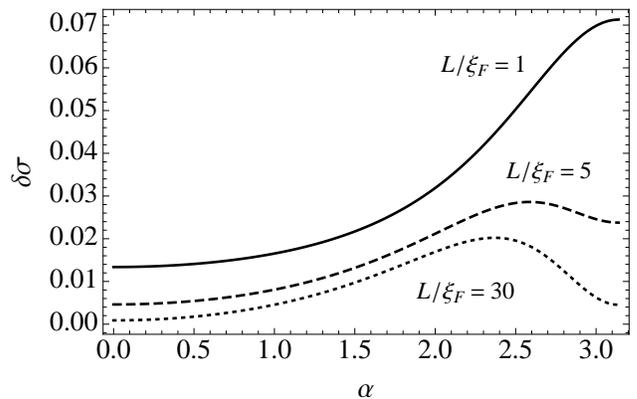}} \caption{Correction to conductance, $\delta\sigma$, vs angle between magnetization
vectors, $\alpha$, in the case of large value of parameter $\gamma_{BS}$
for $V=0$ and $T=0$ at $h_{1}=2$, $h_{2}=5$, for $L/\xi_{F}=1$
(solid line)$,L/\xi_{F}=5$ (dashed line), $L/\xi_{F}=30$ (dotted
line). }
\end{figure}

Fig. 2 shows a dependence of correction to differential conductance
($\delta\sigma=\sigma-1$) normalized to conductance in the normal
state as a function of misorientation angle $\alpha$ calculated
at zero voltage from (\ref{eq:condlin})-(\ref{eq:ff-1}) for three
different lengths of the long ferromagnetic film ($L/\xi_{F}=1,L/\xi_{F}=5,L/\xi_{F}=30$).
Here we consider that exchange energies of the two F films are different
($h_{1}=2,h_{2}=5$). Note that 
at $h_{1}=h_{2}$ the conditions of the linear approximation are violated
at zero voltage.

For short upper ferromagnetic film ($L/\xi_{F}=1$) conductance rises
from $\alpha=0$ to $\alpha=\pi$ monotonically since all correlations
are still present in the structure and magnetic configuration for
$\alpha=\pi$ corresponds to smaller average exchange energy in comparison
with $\alpha=0$. It is also seen from the figure that the shape of
$\delta\sigma$ dependencies begins to change with increase of $L$.
The reason of this transformation is that singlet and 
short-range triplet components begin to decrease very fast deep into
long parts of upper F film in comparison with 
long-range triplet part that decreases slowly. For $L/\xi_{F}=30$
(dotted line in Fig. 2) as well as in a limit of long upper ferromagnetic
film, $L\gg\xi_{F},$ only these long-range triplet correlations can
be taken into account. Expressions (\ref{eq:condlin})-(\ref{eq:ff})
give in this case
\begin{equation}
\sigma=1+\left(\frac{h_{2}\sin(\alpha)/\gamma_{BS}}{h_{1}^{2}+h_{2}^{2}+2h_{1}h_{2}\cos(\alpha)}\right)^{2}.\label{eq:siTR}
\end{equation}
It is seen that in a vicinity of the angles $\alpha=0$ and $\alpha=\pi$
correction to the conductivity is negligible. This is a consequence
of the fact that at these angles there are no long-range triplet correlations
in the structure.

The dependence of differential conductance has a maximum at some intermediate
angle

\[
\alpha_{m}=\arccos(\frac{-2h_{1}h_{2}}{h_{1}^{2}+h_{2}^{2}}).
\]
It is seen that the position of the maximum is the function of exchange
energies only and it is always located at $\alpha>\pi/2$ since exchange
energies are different in this case. In the area near $\alpha_{m}$
long-range triplet correlations declare themselves very strongly.

This can also be seen from Fig.3. It shows dependencies of correction
to differential conductance on applied voltage for long upper ferromagnetic
film $L/\xi_{F}=30$ for three different misorientation angles $\alpha=0.5$,
$\alpha=2.4$ and $\alpha=\pi$. At zero voltage there is a strong
peak in differential conductance at $\alpha=2.4$. It is exactly the
angle at which there is the maximum in $\delta\sigma(\alpha)$ dependence
calculated for $V=0$ (see dotted line in Fig. 2). With inclination
of angle $\alpha$ from $\alpha_{m}$ the height of the peak decreases
since influence of long-range triplet component decreases. 

\begin{figure}
\centerline{\includegraphics{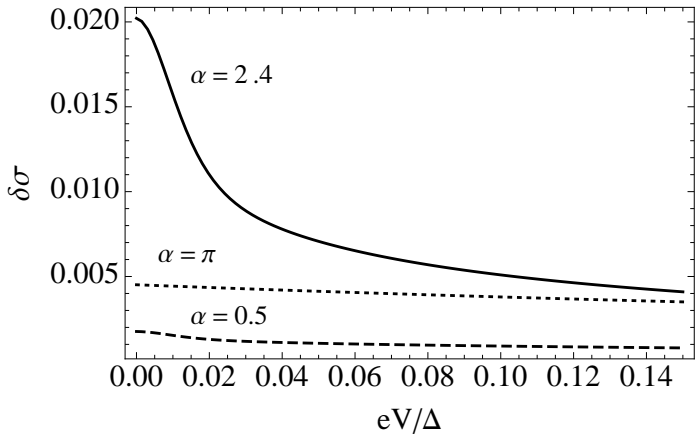}} \caption{Correction to conductance, $\delta\sigma$, vs applied voltage $eV/\Delta$
at $h_{1}=2$, $h_{2}=5$, $L/\xi_{F}=30$ and $T=0$ for $\alpha=0.5$
(dashed line), $\alpha=2.4$ (solid line) and $\alpha=\pi$ (dotted
line).}
\end{figure}

Let us discuss behavior of differential conductance beyond the linearized
case for arbitrary value of parameter $\gamma_{BS}$. To calculate
it, the nonlinear equations (\ref{eq:us_g})-(\ref{eq:bc_2}) were
solved numerically using shooting method.

Fig. 4 shows dependence of conductance of a long upper ferromagnetic
film on angle $\alpha$ at zero voltage. This conductance is caused
by long-range triplet correlations since for long ferromagnetic film
$L/\xi_{F}=30$ only these correlations are strong enough. At $\gamma_{BMS}=0.4$
behavior of conductance is still similar to the one seen from Fig.2.
As SF interface becomes more transparent (suppression parameter $\gamma_{BMS}$
decreases) peaks on the graph get higher. Further reduction of the
suppression parameter results in appearance of an angle interval in
which correction to differential conductance is zero, peaks on the
boarders of the interval increase sharply(see dashed line Fig. 4).

\begin{figure}
\centerline{\includegraphics[scale=0.9]{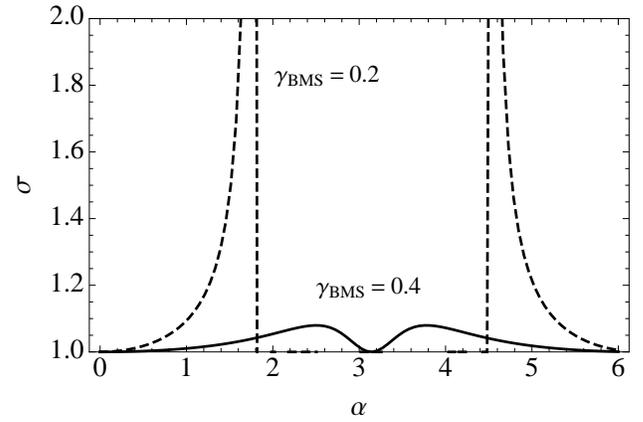}} \caption{Conductance, $\sigma$, vs angle between magnetization vectors, $\alpha$,
at zero voltage, $h_{1}=2$, $h_{2}=5$, $L/\xi_{F}=30$ for two different
values of parameter $\gamma_{BMS}$: $\gamma_{BMS}=0.2$ (dashed line),
$\gamma_{BMS}=0.4$ (solid line).}
\end{figure}

\begin{figure}
\centerline{\includegraphics[scale=0.9]{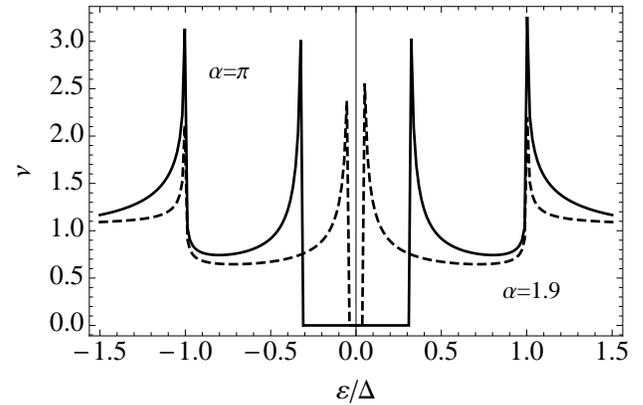}} \caption{Density of states, $\nu$, vs energy at zero voltage, $h_{1}=2$,
$h_{2}=5$, $L/\xi_{F}=30$, $\gamma_{BMS}=0.2$ for two values of
misorientation angle between magnetization vectors $\alpha=\pi$ (solid
line) and $\alpha=1.9$ (dashed line).}
\end{figure}

\begin{figure}
\centerline{\includegraphics[scale=0.9]{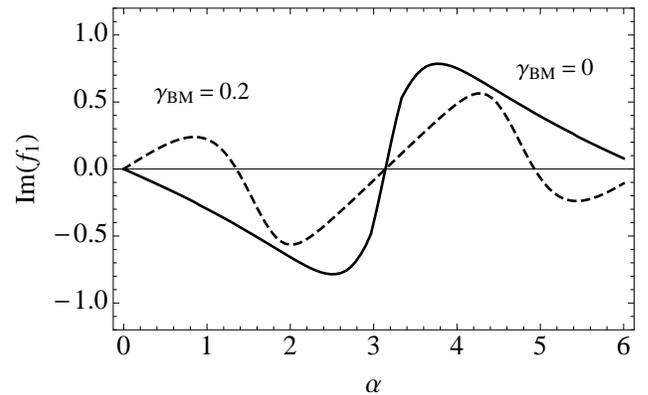}} \caption{Imaginary part of equal spin triplet condensate function, $f_{1}$,
vs angle between magnetization vectors, $\alpha$, for zero voltage
at $h_{1}=2$, $h_{2}=5$, $y=0$, $\gamma_{BMS}=0.4$ for two different
suppression parameters $\gamma_{BM}=0$ (solid line) and $\gamma_{BM}=0.2$
(dashed line). }
\end{figure}

Appearance of strong peaks can be understood from Fig. 5. It shows
dependence of DoS, $\nu=Re(g_{0})$, on energy, $\varepsilon,$ calculated
numerically for $h_{1}=2$, $h_{2}=5$, $L/\xi_{F}=30$, $\gamma_{BMS}=0.2$
and two values of misorientation angle between magnetization vectors
$\alpha=\pi$ (solid line) and $\alpha=1.9$ (dashed line). At $\alpha=\pi$
there are two peaks in density of states located at $\varepsilon=\Delta$
and at the position of minigap. Deviation of the angle $\alpha$ from
$\alpha=\pi$ leads to increase of effective exchange energy. As the
result, the position of minigap shifts to smaller energy and there
exists an angle $\alpha$ at which the minigap becomes zero and position
of the peak in the density of states is localized at zero energy.
It occurs at $\alpha=1.8$ that is exactly at the angle $\alpha,$
at which there are peaks in dependence $\sigma(\alpha)$ presented
in Fig. 4.

As it was discussed earlier in \cite{Karminskaya2} for SF1F2 structure,
suppression parameter at F1F2 interface can strongly influence on
behavior of long-range triplet correlations leading to additional
phase-slip at F1F2 interface. Indeed, taking into account nonzero value
of parameter $\gamma_{BM}$ at F1F2 interface we obtain that behavior
of component $f_{1}$ in F1-SF1F2-F1 structure changes significantly
(Fig. 6). Dashed line ($\gamma_{BM}=0.2$) shows that triplet component
changes it's sign at some intermediate angle, which is not equal to
$0$ or $\pi$. The magnitude of this angle depends on difference
between $h_{1}$ and $h{}_{2}$, and on suppression parameters $\gamma_{BS}$
and $\gamma_{BM}$. With increase of suppression parameters and decrease
of $\mid h_{1}-h_{2}\mid$ it moves towards $\alpha=\pi/2$.

Long-range triplet correlations prevail in the long F1 film of F1-SF1F2-F1
structure, so the features that are seen on Fig. 6 will remain in
conductance. Fig. 7 shows dependence of differential conductance of
upper F1 film on angle $\alpha$ at zero voltage for several values
of suppression parameter $\,\gamma_{BM}=0,0.01,0.2$. With increase
of $\gamma_{BM}$ the shape of conductance changes due to changing
of the long-range triplet component (dashed line and dotted lines).
In the case of finite transparency of F1F2 interface one maximum in
differential conductance transforms into two maximums. Also at $\gamma_{BM}=0.01$
maximum value of conductance (dashed line) can be even larger than
for structure with ideal transparency of F1F2 interface (solid line).
This fact is in good agreement with the discussion performed in \cite{Karminskaya2}.
\begin{figure}
\centerline{\includegraphics[scale=0.9]{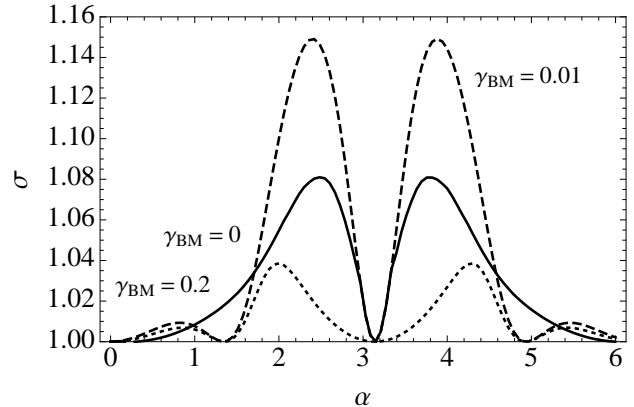}} \caption{Conductance, $\sigma$, vs angle between magnetization vectors, $\alpha$,
for zero voltage at $h_{1}=2$, $h_{2}=5$, $L/\xi_{F}=30$, $\gamma_{BS}=0.4$
for two different suppression parameters $\gamma_{BM}=0$ (solid line),
$\gamma_{BM}=0.01$ (dashed line) and $\gamma_{BM}=0.2$ (dotted line). }
\end{figure}

In conclusion, we have investigated conductance of a long ferromagnetic
film in F1-SF1F2-F1 structure. In the collinear magnetization case, the conductance rapidly decreases
with increase of length of the F1 film. However, in the configuration
with noncollinear magnetizations the conductance decrease slowly due to the generation of long-range triplet
superconducting correlations.
Strong dependence of the differential conductance on
misorientation angle allows to control the conductance by changing directions
of magnetization of one ferromagnetic film.
Further, we demonstrate that long-range triplet correlations manifest themselves as a zero-bias peak in the case of perfect transparency of F1F2 interface,
while a two-peak structure is realized in the case of finite transparency. These features may serve as a
diagnostic tool for characterization of interfaces in superconducting hybrid structures.

The work is supported in part by Russian and Belarusian Funds for Basic Research under RFBR-BFBR
Grants No. l2-02-90010 (M.Yu.K.)and No. Fl2R-014 (S.L.P.), EU-Japan collaboration program "IRON SEA"
and by the Ministry of Education and Science of the Russian Federation.

\end{document}